\documentclass{article}
\usepackage[preprint]{spconfa4}
\usepackage{amsmath,amssymb,graphicx}
\usepackage[utf8]{inputenc}
\usepackage[T1]{fontenc}
\usepackage[english]{babel} 
\usepackage{xcolor}
\usepackage[printonlyused,nolist]{acronym}
\usepackage{multirow}
\usepackage{algorithm}
\usepackage{algpseudocode}
\usepackage{url}
\usepackage{cite}
\usepackage{booktabs} 

\begin{acronym}
	\acro{STFT}{short-time Fourier transform}
	\acro{SOI}{source of interest}
	\acro{ATF}{acoustic transfer function}
	\acro{AEC}{acoustic echo cancellation}
	\acro{DNN}{deep neural network}
	\acro{ENR}{echo-to-noise power ratio}
	\acro{SIR}{SOI-to-interference power ratio}
	\acro{SIER}{SOI-to-interference-plus-echo power ratio}
	\acro{SER}{SOI-to-echo power ratio}
	\acro{ERLE}{echo return loss enhancement}
	\acro{SNR}{signal-to-noise power ratio}
	\acro{PDF}{probability density function}
	\acro{RIR}{room impulse response}
	\acro{IVE}{independent vector extraction}
	\acro{ICE}{independent component extraction}
	\acro{IVA}{independent vector analysis}
	\acro{ICA}{independent component analysis}
	\acro{LS}{least squares}
	\acro{OC}{orthogonality constraint}
	\acro{BSS}{blind source separation}
	\acro{BSE}{blind source extraction}
	\acro{BF}{beamformer}
	\acro{BNLMS}{batch-normalized least mean squares}
\end{acronym}
   
\DeclareMathOperator{\bx}{\mathnormal{\boldsymbol{x}_{f, \tau}}}
\DeclareMathOperator{\bs}{\mathnormal{\boldsymbol{s}_{f, \tau}}}
\DeclareMathOperator{\bd}{\mathnormal{\boldsymbol{d}_{f, \tau}}}
\DeclareMathOperator{\bn}{\mathnormal{\boldsymbol{n}_{f, \tau}}}

\DeclareMathOperator{\be}{\mathnormal{\boldsymbol{e}_{f, \tau}}}
\DeclareMathOperator{\beH}{\mathnormal{\boldsymbol{e}_{f, \tau}^{\text{H}}}}
\DeclareMathOperator{\bu}{\mathnormal{{u}_{f, \tau}}}
\DeclareMathOperator{\bq}{\mathnormal{\boldsymbol{q}_{f, \tau}}}
\DeclareMathOperator{\sHat}{\mathnormal{\hat{{s}}_{f, \tau}}}
\DeclareMathOperator{\bsHatBroad}{\mathnormal{\hat{\underline{\boldsymbol{s}}}_{\tau}}}
\DeclareMathOperator{\bzHat}{\mathnormal{\hat{\boldsymbol{z}}_{f, \tau}}}
\DeclareMathOperator{\bzHatBroad}{\mathnormal{\hat{\underline{\boldsymbol{z}}}_{\tau}}}

\DeclareMathOperator{\bH}{\mathnormal{\boldsymbol{h}_{f}}}

\DeclareMathOperator{\bwBse}{\mathnormal{\boldsymbol{w}_{\text{bse},f}}}
\DeclareMathOperator{\bwBseH}{\mathnormal{\boldsymbol{w}_{\text{bse},f}^{\text{H}}}}

\DeclareMathOperator{\bg}{\mathnormal{\boldsymbol{g}_{f}}}
\DeclareMathOperator{\baSoi}{\mathnormal{\boldsymbol{a}_{\text{soi},f} }}
\DeclareMathOperator{\bABg}{\mathnormal{\boldsymbol{A}_{bg,f}}}
\DeclareMathOperator{\bB}{\mathnormal{\boldsymbol{B}_{f}}}
\DeclareMathOperator{\bBH}{\mathnormal{\boldsymbol{B}_{f}^{\text{H}}}}
\DeclareMathOperator{\bHbg}{\mathnormal{\boldsymbol{h}_{\text{bg},f}}}
\DeclareMathOperator{\gam}{\mathnormal{\gamma_f}}

\DeclareMathOperator{\phiNl}{\mathnormal{\phi_f}}
\DeclareMathOperator{\numMics}{\mathnormal{M}}
 
\newcommand{\commentTHa}[1]{\textcolor{black}{#1}}
                
\title{Joint Acoustic Echo Cancellation and Blind Source Extraction \\based on Independent Vector Extraction}
\twoauthors
  {\hspace*{-0.3cm}Thomas Haubner and Walter Kellermann}
	{\hspace*{-0.3cm}Multimedia Communications and Signal Processing,\\
	\hspace*{-0.3cm}Friedrich-Alexander-Universit\"at Erlangen-N\"urnberg,\\
	\hspace*{-0.3cm}Cauerstr. 7, D-91058 Erlangen}
    {\hspace*{-1.75cm}Zbyněk Koldovský\thanks{This work was partly supported by The Czech Science Foundation through Project No.~20-17720S.}}
	{\hspace*{-1.75cm}Acoustic Signal Analysis and Processing Group,\\
	\hspace*{-1.75cm}Technical University of Liberec,\\
	\hspace*{-1.75cm}Studentská 2, CZ-46117 Liberec}

\begin{document}
%
%
\maketitle
\fontsize{9.5pt}{11.5pt}\selectfont
\begin{abstract}
We describe a joint acoustic echo cancellation (AEC) and blind source extraction (BSE) approach for multi-microphone acoustic frontends. The proposed algorithm blindly estimates AEC and beamforming filters by maximizing the statistical independence of a non-Gaussian source of interest and a stationary Gaussian background modeling interfering signals and residual echo. Double talk-robust and fast-converging parameter updates are derived from a global maximum-likelihood objective function resulting in a computationally efficient Newton-type update rule. Evaluation with simulated acoustic data confirms the benefit of the proposed joint AEC and beamforming filter estimation in comparison to updating both filters individually.\let\thefootnote\relax\footnotetext{Accepted for \textit{Int. Workshop on Acoust. Signal Enhanc. (IWAENC 2022)}.} 
\end{abstract}
\section{Introduction}
\label{sec:intro}
Acoustic echo and interference suppression is a crucial part of any modern hands-free speech communication device \cite{9413457}. It has been tackled by a variety of approaches ranging from traditional adaptive linear filters to sophisticated deep learning-based spectral post filters and beamformers \cite{Kellermann1997StrategiesFC, haensler2004acoustic,ENZNER20061140, 9413457, luis_valero_2019, 9414868}. In particular, the combination of \ac{AEC} and beamforming algorithms have proven to be a powerful approach due to its simultaneous exploitation of spectral and spatial signal characteristics \cite{Kellermann1997StrategiesFC,kellermann_acoustic_2001}. Yet, sophisticated parameter estimation schemes are required to cope with time-varying echo paths, high-level double-talk and interference that many devices are exposed to \cite{Enzner2014AcousticEC}. This problem has been addressed by Kalman filter-based inference of the model parameters \cite{ENZNER20061140,park_state-space_2019, cohen_online_2021} and machine-learning supported variants \cite{9616295, 9414180, 9747334}. Besides Kalman filter-based approaches, also \textit{blind} algorithms originating from \ac{ICA} (see, e.g., \cite{comon1994independent}) have shown promising results for interference-robust adaptation control \cite{nesta_batch_online, gunther_learning_2012,9414623, 9357975}. They are particularly interesting as they require no prior training and thus are very robust w.r.t. training-testing mismatches which any machine learning-based approach suffers from. So far, semi-supervised \ac{BSS} has been mainly used for double-talk robust adaptation control of linear and nonlinear acoustic echo cancellers using various statistical source models and update rules \cite{nesta_batch_online, gunther_learning_2012, 9414623, 9357975,4538677}. Recently, also \ac{IVA}-based joint estimation of \ac{AEC} filters and beamforming filters has been investigated \cite{na21_interspeech}. \ac{IVA}-based approaches are promising as they have shown competitive multichannel speech enhancement performance when compared to sophisticated \ac{DNN} algorithms \cite{9657518}. Yet, \cite{na21_interspeech} assumes that the number of microphones $M$ is equal to the number of  statistically independent point sources. This assumption requires to first estimate $M$ beamforming vectors and subsequently select the desired \ac{SOI}, which becomes computationally complex for large $M$. 

Here, we propose a joint \ac{AEC} and \ac{BSE} approach which estimates only a single desired \ac{SOI} out of the observed mixture. For this, we model the frequency-domain \ac{SOI} as a vector source following a multivariate non-Gaussian source \ac{PDF} which is independent from the interfering signals \cite{grad_alg_quest_conv}. The interference is modeled by a stationary and spatially-correlated circular complex Gaussian \cite{grad_alg_quest_conv}. Instead of aiming at a complete separation of the \ac{SOI}, the loudspeakers signals, and the interferers, we allow the interference estimates to remain mutually mixed \cite{grad_alg_quest_conv}. This significantly reduces the number of model parameters to be estimated for large numbers of microphones. The parameters are estimated by a fast-converging Newton-type update which can be interpreted as a joint \ac{SOI}- and interference-aware \ac{BNLMS} echo canceller with a fast independent vector extraction (IVE)-controlled beamformer.

We represent vectors by bold lower-case letters and matrices by bold upper-case letters with $[\cdot]_{m}$ denoting the $m$th element of a vector. The $M$-dimensional identity matrix and the $M\times N$-dimensional zero matrix are denoted by $\boldsymbol{I}_M$ and $\boldsymbol{0}_{M \times N}$, respectively. Finally, we indicate sampling from a \ac{PDF} $p(\boldsymbol{z})$ by $\boldsymbol{z}\sim p(\boldsymbol{z})$, equivalency up to a constant by $\stackrel{\text{c}}{=}$, and averaging over time by $\widehat{\mathbb{E}}$.

\section{Signal Model}
\label{sec:signal_model}
%
We consider a multi-microphone acoustic echo and interference control scenario comprising a single \ac{SOI}, e.g., a desired near-end speaker, a single-channel far-end loudspeaker signal, and various point interferers. The $M$-channel \ac{STFT}-domain microphone signal $\bx$ at frequency index~$f$ and frame index $\tau$ is modeled as a linear superposition of the \ac{SOI} image $\bs$, the echo image $\bd$, and the interference image $\bn$ as follows:
\begin{align}
	\bx  &=  \bs +\bd + \bn  \in \mathbb{C}^M.
	\label{eq:add_sig_mod}
\end{align}
We assume a linear time-invariant multiplicative narrowband transfer function model to describe the acoustic propagation from the \ac{SOI} and the loudspeaker to the microphones
\begin{equation}
	\bs = \baSoi s_{f,\tau} \quad \text{and} \quad \bd = \bH \bu,
	\label{eq:sig_image_mod_1}
\end{equation}
with the \acp{ATF} $\baSoi$ and $\bH$ and the scalar desired source signal \makebox{$s_{f,\tau}$} and the scalar far-end loudspeaker signal \makebox{$\bu$}. Furthermore, we assume the interfering signal image $\bn$ to be composed of $M-1$ point sources $[\bq]_1$, $\dots$, $[\bq]_{M-1}$, whose acoustic propagation to the microphones is modeled by
\begin{equation}
	\bn = \bABg \bq
	\label{eq:interference_mod}
\end{equation}
with the mixing matrix \makebox{$\bABg \in \mathbb{C}^{M \times M-1}$}.
%
Note that this assumption expresses a determined mixing model, i.e., an equal number of observations and latent point sources, which allows for a detailed model analysis in Sec. \ref{sec:dem_model}.

\section{Joint Acoustic Echo Cancellation and Blind Source Extraction}
\label{sec:joint_aec_ive}
We will now describe the proposed joint \ac{AEC} and \ac{BSE} algorithm by first introducing the proposed demixing model and then the blind estimation of its coefficients by maximizing the mutual statistical independence of the \ac{SOI} estimate, the interference estimate and the loudspeaker signal.

\subsection{Demixing Model}
\label{sec:dem_model}
%
To estimate the \ac{SOI} $s_{f,\tau}$, we use the semi-supervised demixing model \cite{nesta_batch_online}
\begin{align}
	\begin{pmatrix}
		\sHat \\ \bzHat \\ \bu
	\end{pmatrix}& = {\boldsymbol{W}}_f 	\begin{pmatrix}
	\bx \\ \bu
	\end{pmatrix} 
	\label{eq:dem_model}
\end{align}
with the \ac{SOI} estimate $\sHat$, the background estimate $\bzHat$, modelling interference and residual echo, and the demixing matrix
		\begin{align}
 {\boldsymbol{W}}_f 
	 = 	\begin{pmatrix}
		\bwBseH &   w_{\text{aec},f}^* \\
		\bB & \bHbg \\
		\boldsymbol{0}_{1 \times M} &  1
	\end{pmatrix} \in \mathbb{C}^{\commentTHa{M+1 \times M+1}}
\label{eq:dem_mat}
\end{align}
which is composed of the parameters \makebox{$\bwBse \in \mathbb{C}^M$}, \makebox{$w_{\text{aec},f}\in \mathbb{C}$}, \makebox{$\bHbg \in \mathbb{C}^{M-1}$} and \makebox{$\bB \in \mathbb{C}^{M-1 \times M}$}. 
%
We analyze the signal estimation problem by computing the overall transmission model
\begin{align}
	\begin{pmatrix}
		\sHat \\ \bzHat \\ \bu
	\end{pmatrix}= \boldsymbol{V}_f
	\begin{pmatrix}
		s_{f,\tau} \\ \bq \\ \bu
	\end{pmatrix}
	\label{eq:overal_trans_mod}
\end{align}
between the source signals $s_{f,\tau}$, $\bq$ and $\bu$ and the signal estimates $\sHat$, $\bzHat$ and $\bu$ with the transmission matrix
\begin{align}
	\boldsymbol{V}_f=
	\begin{pmatrix}
		\bwBseH \baSoi & \bwBseH \boldsymbol{A}_{bg,f} & \bwBseH \bH + {w}_{\text{aec},f}^* \\
		\bB \baSoi & \bB \boldsymbol{A}_{bg,f} & \bB \bH + \bHbg \\
		0 & \boldsymbol{0}_{1\times M-1} & 1
	\end{pmatrix}.
	\label{eq:overal_trans_matrix}
\end{align}
%
As we aim in \ac{BSE} at the separation of the \ac{SOI} estimate $\sHat$ \commentTHa{from the multivariate background estimate $\bzHat$, with potentially still statistically dependent components,} and the loudspeaker signal $\bu$, we conclude that the optimum demixing matrix \eqref{eq:dem_mat} must enforce a block-diagonal transmission matrix $\boldsymbol{V}_f$.
Note that a complete diagonalization of the transmission matrix is not necessary as we do not require the background estimates $\bzHat$ to be equal to the interfering source signals $\bq$. The block-diagonality assumption allows to couple the different components of the demixing matrix $\boldsymbol{W}_f$ and the mixing model parameters $\baSoi$, $\bABg$ and $\bH$ (cf. Sec.~\ref{sec:signal_model}). In particular, we require that
\begin{equation}
	\bwBseH \baSoi = 1  
	\label{eq:soi_soi_constraint}
\end{equation}
which is well-known as distortionless response constraint by identifying $\bwBse$ as beamforming vector \cite{van_trees}. Furthermore, the \ac{SOI} should not be contained in the background estimate $\bzHat$ which is equivalent to
\begin{equation}
	\bB \boldsymbol{a}_{\text{soi},f}  = \boldsymbol{0}_{\commentTHa{M-1} \times 1}
	\label{eq:bg_soi_const}
\end{equation}
with $\bB$ being identified as blocking matrix. A straightforward choice for $\bB$ is given by \cite{grad_alg_quest_conv}
\begin{equation}
	\bB = \begin{pmatrix} \bg
		& -\gam \boldsymbol{I}_{\commentTHa{M-1}}
	\end{pmatrix}
	\label{eq:block_mat_def}
\end{equation}
which allows to parametrize the blocking matrix $\bB$ entirely in terms of the \ac{SOI} \ac{ATF} $\baSoi = \begin{pmatrix}\gam & \boldsymbol{g}_{f}^{\commentTHa{\text{T}}}\end{pmatrix}^{\text{T}}$. Finally, to ensure that the echo signal is neither contained in the \ac{SOI} estimate $\sHat$ nor in the background estimate $\bzHat$, we must ensure that
\begin{align}
	  w_{\text{aec},f}^*&= -\bwBseH  \bH \label{eq:soi_aec_constraint} \\
	  \bHbg &= -\bB \bH.	\label{eq:bg_aec_constraint}
\end{align}
Note that \eqref{eq:soi_aec_constraint} readily identifies $-w_{\text{aec},f}^*$ as concatenated system of the echo \ac{ATF} $\bH$ and the beamforming vector $\bwBse$.
By replacing $ w_{\text{aec},f}$, $\bHbg$ and $\bB$ (cf.~Eqs.~\eqref{eq:block_mat_def}-\eqref{eq:bg_aec_constraint}) in the upper part of the demixing model \eqref{eq:dem_model}, we obtain
\begin{align}
	\begin{pmatrix}
		\sHat \\ \bzHat
	\end{pmatrix}& =
 	\begin{pmatrix}
	\bwBseH  \\
	\begin{pmatrix} \bg
		& -\gam \boldsymbol{I}_{\commentTHa{M-1}}
	\end{pmatrix} \\
\end{pmatrix} 
		\left( \bx - \bH \bu \right)
	\label{eq:dem_model_2}
\end{align}
with the signal estimator being solely parametrized by $\bwBse$, $\baSoi = \begin{pmatrix}\gam & \boldsymbol{g}_{f}^{\commentTHa{\text{T}}}\end{pmatrix}^{\text{T}}$ and $\bH$. We conclude that the considered constraints lead naturally to the traditional approach of first subtracting the echo estimate $\hat{\boldsymbol{d}}_{f \tau} =  \bH \bu$ from the microphone signal $\bx$ and subsequently applying a beamformer \cite{Kellermann1997StrategiesFC}. In the next sections, we will show how to jointly estimate the unknown parameters $\bwBse$, $\baSoi$ and $\bH$. 

\subsection{Cost Function}
\label{sec:cost_fct}
Adopting the idea of \ac{IVE} \cite{grad_alg_quest_conv}, we suggest to estimate the parameter vector $\boldsymbol{\theta}$, containing $\bwBse$, $\baSoi$ and $\bH$, by maximizing the statistical independence of the broadband \ac{SOI} and background signal estimates {$ \hat{\underline{\boldsymbol{s}}}_\tau^{\text{T}} = \begin{pmatrix}	\hat{s}_{1,\tau} & \dots & \hat{s}_{F,\tau}	\end{pmatrix} $} and $ \hat{\underline{\boldsymbol{z}}}_\tau^{\text{T}}= \commentTHa{\begin{pmatrix}	\hat{\boldsymbol{z}}_{1,\tau}^{\text{T}} & \dots & \hat{\boldsymbol{z}}_{F,\tau}^{\text{T}}	\end{pmatrix}}$. The according  log-likelihood function is given by \cite{grad_alg_quest_conv}
\begin{align}
\mathcal{L}(\boldsymbol{\theta})
\stackrel{\text{c}}{=}  \sum_{\tau=1}^T &\left( \log p  (\bsHatBroad) + \log p (\bzHatBroad )  \right) + 2 T \sum_{f=1}^F\log \left| \det {\boldsymbol{W}}_f  \right| 
\label{eq:ll_fct}
\end{align}
with the number of frames $T$, the number of frequency bins $F$, and the true signal models \makebox{$\underline{\boldsymbol{s}}_{\tau} \sim p  (\underline{\boldsymbol{s}}_{\tau})$} and \makebox{$\underline{\boldsymbol{z}}_{\tau} \sim p(\underline{\boldsymbol{z}}_{\tau} )$}, respectively. Dividing \eqref{eq:ll_fct} by $-T$ and assuming the background signal $\underline{\boldsymbol{z}}_{\tau}$ to follow a stationary zero-mean circular complex Gaussian \ac{PDF} with mutually uncorrelated frequency bins, i.e., $p(\underline{\boldsymbol{z}}_{ \tau})= \prod_{f=1}^{F} \mathcal{N}_c(\boldsymbol{0}_{M-1\times1}, \boldsymbol{C}_{{z} {z},f})$, we obtain the cost function: 
\begin{align}
	\begin{split}
\mathcal{J}(\boldsymbol{\theta}) \stackrel{\text{c}}{=} & ~\mathbb{\widehat{E}}\left[ \commentTHa{-} \log p (\bsHatBroad ) + \sum_{f=1}^F   \beH  \boldsymbol{R}_f \be  \right]  \\&\qquad\qquad\qquad\qquad- (\numMics-2) \sum_{f=1}^F \log \left|   \gam\right|^{2} 
\end{split}
\label{eq:cost_func}
\end{align}
with the \ac{AEC} error signal $\be =\bx - \bH \bu $, the first component of the \ac{SOI} \ac{ATF} $\gamma_f = \left[ \baSoi\right]_1$  and {$\boldsymbol{R}_f =\bBH \boldsymbol{C}_{{{z}{z},f}}^{-1}  \bB $} \cite{grad_alg_quest_conv}. It has been observed that $\bwBse$ and $\baSoi$ are often not sufficiently strongly coupled by the cost function \eqref{eq:cost_func} \cite{grad_alg_quest_conv}. This shortcoming is addressed by the \ac{OC} $\widehat{\mathbb{E}} \left[ \bzHat \hat{s}_{f,\tau}^* \right] = \boldsymbol{0}_{M-1\times 1}$ \cite{cois1994performance}, which leads to the relation \cite{grad_alg_quest_conv}
\begin{equation}
	\baSoi = \begin{pmatrix}
		\gam \\ \bg
	\end{pmatrix} = \frac{\widehat{\boldsymbol{C}}_{ee,f}\bwBse }{\bwBseH\widehat{\boldsymbol{C}}_{ee,f} \bwBse}
	\label{eq:orth_constraint}
\end{equation}
between the \ac{SOI} \ac{ATF} $\baSoi$ and the beamforming vector $\bwBse$ with $\widehat{\boldsymbol{C}}_{ee,f} = \widehat{\mathbb{E}}\left[ \be \beH \right]$ being the sample error covariance matrix. Note that analogously to \cite{grad_alg_quest_conv}, the \ac{OC} \eqref{eq:orth_constraint} is equivalent to computing $\bwBse$ as the minimizer of $\bwBseH \widehat{\boldsymbol{C}}_{ee,f} \bwBse$ subject to $\bwBseH \baSoi=1$, i.e., a minimum \textit{error} power beamformer which is distortionless w.r.t. $\baSoi$ \cite{8081389}. 
By enforcing the \ac{OC} \eqref{eq:orth_constraint}, the demixing matrix $\boldsymbol{W}_f$ is entirely parametrized by $\bwBse$ and $\bH$ which need to be estimated by minimizing \eqref{eq:cost_func}.

\subsection{Parameter Update}
\label{sec:par_update}
For rapid convergence and optimum steady-state performance, we derive a Newton-type update (see, e.g., \cite{complex_ad_filt}) with a block-diagonal Hessian, i.e., $\frac{\partial^2 \mathcal{J}(\boldsymbol{\theta}) }{\partial \boldsymbol{w}_{\text{bse},f} \partial \boldsymbol{h}_{f}^{\text{H}}  }=\frac{\partial^2 \mathcal{J}(\boldsymbol{\theta}) }{\partial \boldsymbol{w}_{\text{bse},f} \partial \boldsymbol{h}_{f}^{\text{T}}  }= \boldsymbol{0}_{M \times M} $, to estimate $\bwBse$ and $\bH$. The first-order partial derivatives are given by
\begin{align}
\frac{\partial  \mathcal{J}(\boldsymbol{\theta}) }{\partial \boldsymbol{h}_{f}^*} & =	-\mathbb{\widehat{E}} \left[   \left( \phi_f^*(\bsHatBroad)   \bwBse 
+ \boldsymbol{R}_f  \be  \right) {u}_{f\commentTHa{,} \tau}^* \right] \label{eq:grad_haec}  \\
\frac{\partial \mathcal{J}(\boldsymbol{\theta}) }{\partial \boldsymbol{w}_{\text{bse},f}^*}   &=  \mathbb{\widehat{E}} \left[ \be \phiNl(\bsHatBroad) \right] - \baSoi 
\label{eq:grad_wbse}
\end{align}
with the score function $\phiNl(\bsHatBroad) = - \frac{\partial \log \commentTHa{p (\bsHatBroad)}}{\partial \hat{s}_{f, \tau} }$ and the \ac{SOI} \ac{ATF} $\baSoi$ given by \eqref{eq:orth_constraint}. We conclude that \eqref{eq:grad_wbse} corresponds to the \ac{BSE} filter gradient computed in \cite{grad_alg_quest_conv,9387552} when exchanging the \ac{AEC} error signal $\be$ by the microphone signal $\bx$ (see also sequential demixing model \eqref{eq:dem_model_2}). In \cite{grad_alg_quest_conv,9387552} it is shown that the score function must fulfill \makebox{$\mathbb{E}\left[ s_{f,\tau} \phiNl(\underline{\boldsymbol{s}}_{f,\tau}) \right]=1$} to obtain a valid update, which can be ensured by normalizing $\phiNl(\bsHatBroad)$ in Eqs. \eqref{eq:grad_haec} and \eqref{eq:grad_wbse} with $\hat{\nu}_f = \mathbb{\widehat{E}} \left[ \sHat \phiNl(\bsHatBroad)  \right]$. By following analogous steps as in \cite{9387552} we obtain the Newton-type \ac{BSE} filter update
%
\begin{align}
		\bwBse \gets \bwBse +  \frac{\hat{\nu}_f^* }{ \hat{\rho}_f^* - \hat{\nu}_f^* }  \widehat{\boldsymbol{C}}_{ee,f}^{-1} 	\left( \mathbb{\widehat{E}} \left[ \be \frac{\phiNl(\bsHatBroad)}{\hat{\nu}_f}  \right] - \baSoi \right)
	\label{eq:newton_update_bse}
\end{align}
%
%
with $\hat{\rho}_f=\mathbb{\widehat{E}} \left[ \frac{\partial \phiNl(\bsHatBroad)  }{\partial \hat{s}_{f,\tau}^*   } \right]$. Note that the update \eqref{eq:newton_update_bse} corresponds to the well-known one-unit FastICA/FastIVA algorithms \cite{fica, 9387552} applied to the \ac{AEC} error signal $\be$. We continue by computing the second-order derivatives for the \ac{AEC} filter update 
\begin{align}
&	\frac{\partial^2 \mathcal{J}(\boldsymbol{\theta}) }{\partial \bH \partial \boldsymbol{h}_{f}^{\text{H}}  } =  \left( \boldsymbol{R}_f^{\text{T}} + \hat{\rho}_f \boldsymbol{w}_{\text{bse},f}^* \boldsymbol{w}_{\text{bse},f}^{\text{T}} \right) \widehat{\mathbb{E}} \left[ \left| \bu \right|^2 \right] \label{eq:sec_der_aec_filt}\\ 
&	\frac{\partial^2 \mathcal{J}(\boldsymbol{\theta}) }{\partial \bH \partial \boldsymbol{h}_{f}^{\text{T}}  } \hspace*{.05cm} =    \hat{\xi}_f   \boldsymbol{w}_{\text{bse},f}^* \boldsymbol{w}_{\text{bse},f}^{\text{H}}  \widehat{\mathbb{E}} \left[   u_{f,\tau}^2 \right] \label{eq:pseudo_hessian_of_aec_filt} 
\end{align}
with $\hat{\xi}_f=\mathbb{\widehat{E}} \left[ \frac{\partial \phiNl(\bsHatBroad)  }{\partial \hat{s}_{f,\tau}   } \right] $.
As most common \ac{STFT}-domain loudspeaker signals, e.g., speech and music, can be well-modeled as circular random variables, $\widehat{\mathbb{E}} \left[   u_{f,\tau}^2 \right]$ will be close to zero, and thus Eq.~\eqref{eq:pseudo_hessian_of_aec_filt} is approximately equal to the all-zero matrix $\boldsymbol{0}_{M \times M}$. This simplifies the Newton step to \cite{complex_ad_filt}
\begin{align}
		\bH \gets \bH - \left(\left(\frac{\partial^2 \mathcal{J}(\boldsymbol{\theta}) }{\partial \bH \partial \boldsymbol{h}_{f}^{\text{H}}  }  \right)^* \right)^{-1} \frac{\partial  \mathcal{J}(\boldsymbol{\theta}) }{\partial \boldsymbol{h}_{f}^*}. 	\label{eq:newton_update_gen_aec} 
\end{align}
By incorporating the score function normalization into \eqref{eq:grad_haec} and \eqref{eq:sec_der_aec_filt} we obtain the proposed \ac{AEC} filter update
\begin{align}
	 \bH &\gets \bH +\left( \left( \boldsymbol{R}_f + \frac{\hat{\rho}_f^{\commentTHa{*}} }{\hat{\nu}_f^{\commentTHa{*}}  }  \boldsymbol{w}_{\text{bse},f} \boldsymbol{w}_{\text{bse},f}^{\text{H}} \right) \widehat{\mathbb{E}} \left[ \left| \bu \right|^2 \right]  \right)^{-1} \nonumber \\ & \hspace*{1.75cm}	\mathbb{\widehat{E}} \left[   \left( \frac{\phi_f^*(\bsHatBroad)}{ \hat{\nu}_f^* }    \bwBse 
+ \boldsymbol{R}_f  \be  \right) {u}_{f \commentTHa{,} \tau}^* \right].
\label{eq:prop_newton_update_aec}
\end{align}
%
The \ac{AEC} filter update \eqref{eq:prop_newton_update_aec} can be interpreted as \commentTHa{an} \ac{SOI}- and interference-aware multichannel \ac{BNLMS}. The traditional single-channel \ac{BNLMS} is obtained by setting the number of microphones $M=1$, which simplifies the beamforming vector to $\commentTHa{\mathnormal{{w}_{\text{bse},f}}} =1$, i.e., $\sHat=e_{f,\tau}$, and discards $\boldsymbol{R}_f$, and choosing a stationary Gaussian \ac{SOI} source model with score function $\phiNl(\bsHatBroad) = \hat{s}_{f,\tau}^*$ which leads to
\begin{align}
 \frac{\partial  \mathcal{J}(\boldsymbol{\theta}) }{\partial {h}_{f}^*}	=	-\mathbb{\widehat{E}} \left[     e_{f,\tau}    {u}_{f \commentTHa{,} \tau}^* \right] \hspace*{.19cm} \text{and} \hspace*{.23cm}
\frac{\partial^2 \mathcal{J}(\boldsymbol{\theta}) }{\partial h_f \partial {h}_{f}^*  } = \mathbb{\widehat{E}} \left[ \left| \bu \right|^2 \right] .
\label{eq:bnlms}
\end{align}
The proposed \ac{AEC} update extends the \ac{BNLMS} (cf.~Eqs. \eqref{eq:newton_update_gen_aec} and \eqref{eq:bnlms}) by incorporating knowledge about the spatial and spectral signal characteristics of the \ac{SOI} and interference and thus acts as an inherent adaptation control.

\subsection{Algorithmic Description}
\label{sec:alg_descr}
The proposed joint \ac{AEC} and \ac{BSE} algorithm  is summarized in Alg.~\ref{alg:prop_alg_descr}.
Thereby we approximate the covariance matrix ${\boldsymbol{C}}_{{{z}{z}},f}$ in $\boldsymbol{R}_f =\bBH \boldsymbol{C}_{{{z}{z}},f}^{-1}  \bB $ by the estimate \makebox{$\widehat{\boldsymbol{C}}_{{{z}{z}},f} = \widehat{\mathbb{E}}\left[ \hat{\boldsymbol{z}}_{f,\tau} \hat{\boldsymbol{z}}_{f,\tau}^{\text{H}} \right]$} and normalize the $\ac{BSE}$ filter $\bwBse$ after each iteration by $\bwBseH \widehat{\boldsymbol{C}}_{ee,f}  \bwBse $ (cf.~line~4) to ensure that the \ac{SOI} estimate $\sHat$ has unit scale, i.e., \makebox{$\mathbb{\widehat{E}}\left[\left| \sHat \right|^2 \right]=1$} \cite{grad_alg_quest_conv}. 
Finally, to address the scale ambiguity, which is inherent to any \ac{ICA} approach \cite{comon1994independent}, we multiply the \ac{SOI} estimate $\sHat$ after the optmization by \makebox{$\min_{\alpha_f} \mathbb{\widehat{E}}\left[ \left| \alpha_f \sHat - [\boldsymbol{e}_{f,\tau}]_r \right|^2 \right]$} \cite{matsuoka2002minimal} which corresponds to a projection of $\sHat$ onto the $r$th error signal $[\boldsymbol{e}_{f,\tau}]_r$. Note that, because of the reduced echo power, we use the error signal as reference instead of the commonly used microphone signal.
%
%
%
%
\begin{figure}[t]
\vspace*{-.65cm}
\begin{algorithm}[H]
	  \caption[Algorithmic description]{Algorithmic description of the proposed joint \ac{AEC} and \ac{BSE} algorithm\footnotemark.}
	\label{alg:prop_alg_descr}
	\begin{algorithmic}[1]
	\For{each iteration}
	    \State Update \ac{AEC} filter $\bH$ by Eq.~\eqref{eq:prop_newton_update_aec}
		\State Update \ac{BSE} filter: $\bwBse$ by Eq.~\eqref{eq:newton_update_bse}
		\State Normalize: $\bwBse \gets \bwBse / (\bwBseH \widehat{\boldsymbol{C}}_{ee,f}  \bwBse) $
	\EndFor
	\State Backprojection: $\sHat \gets  \mathbb{\widehat{E}}[ \hat{s}_{f,\tau}^* [\be]_r  ] / \mathbb{\widehat{E}}[ |\sHat |^2 ]   \sHat $
	\end{algorithmic}
	\vspace*{0.0cm}
\end{algorithm}
	\vspace*{-0.5cm}
\end{figure}
\footnotetext{Source code implementation will be made available at \textcolor{black}{\url{https://github.com/ThomasHaubner/joint_AEC_BSE}}}

\section{Experimental Evaluation}
\label{sec:exp_eval}
The proposed joint \ac{AEC} and \ac{BSE} algorithm is evaluated for a variety of challenging multi-microphone acoustic echo and interference cancellation scenarios. We consider a $4$-element circular microphone array with random diameter in the range $[5\text{cm},~15\text{cm}]$ which is placed randomly in a shoebox room with dimensions in the ranges $[4\text{m},~6\text{m}]$, $[4\text{m},~6\text{m}]$ and $[2.2\text{m},~3.2\text{m}]$, respectively, with random reverberation time $T_{60}\in [0.2\text{s},~0.3\text{s}]$. The loudspeaker, \ac{SOI} and a single interferer are located orthogonally to the microphone array axis with random angles of arrival and random distances in the ranges $[5\text{cm},~20\text{cm}]$, $[0.7\text{m},~1.4\text{m}]$ and $[1.7\text{m},~2.5\text{m}]$, respectively. All \commentTHa{\aclp{RIR}} are simulated according to the image method \cite{habets2010room} with a minimum filter length of $\text{max}(4000, \lceil f_s T_{60}  \rceil )$ and sampling frequency $f_s=16 \text{kHz}$. The loudspeaker, \ac{SOI} and interferer emit $5\text{s}$ speech signals which are sampled randomly from a subset of the LibriSpeech corpus \cite{libri_speech} including $100$ speakers. The \ac{SOI} and interfering signal images (cf.~Eq.~\eqref{eq:add_sig_mod}) are scaled according to a random \ac{SOI}-to-echo and interference-to-echo ratio in the ranges $[5 \text{dB}, ~10\text{dB}]$ and $[0 \text{dB}, ~5\text{dB}]$, respectively. \commentTHa{Note that we consider a dominant \ac{SOI} to alleviate the outer permutation problem, i.e., selecting the correct \ac{SOI}. To simulate} more realistic scenarios, we added spatially uncorrelated white Gaussian \commentTHa{sensor} noise with a random echo-to-noise ratio in the range $[25 \text{dB}, ~35\text{dB}]$.

We chose an \ac{STFT} frame length and frameshift of $2048$ and $1024$, respectively. The \commentTHa{filter vectors} $\bH$ and $ \bwBse$ were updated by $50$ iterations after being initialized with {$\bH=\boldsymbol{0}_{M \times 1}$ and \makebox{$\bwBse=\begin{pmatrix} 1& \boldsymbol{0}_{1 \times M-1}\end{pmatrix}^{\text{T}}$}}, respectively. 
%
%
%
As score function we considered
\begin{equation}
 \phi_{f}(\bsHatBroad) = \frac{\hat{s}_{f,\tau}^* }{ \sqrt{\sum_{f=1}^{F} |\hat{s}_{f,\tau}|^2   }}
\end{equation}
which models joint broadband activity of the \ac{SOI}. We backprojected the \ac{SOI} estimates (cf.~Sec.~\ref{sec:alg_descr}) to the first error signal.

As performance measures we use the logarithmic time-domain \ac{SIR}, \ac{SER}, \ac{SIER} and the logarithmic \ac{ERLE}  \cite{Enzner2014AcousticEC} after the \ac{AEC} unit and the beamformer, respectively. All performance measures are averaged over $50$ experiments with randomly drawn speech signals and randomly sampled acoustic scenarios, i.e., sampling room geometry and reverberation time, microphone array properties, and positions of \ac{SOI}, interferer, and loudspeaker, respectively, in the ranges given above. 

Tab.~\ref{tab:perfEvalTab} shows the average performance of the unprocessed microphone signals, a \ac{LS} \ac{AEC} filter estimate, a fast \acs{IVE}-controlled beamformer, i.e., discarding the \ac{AEC} unit, and finally the proposed joint \ac{AEC}+\acs{IVE} in comparison to the individual update (BNLMS+IVE), i.e., replacing the \ac{AEC} filter update \eqref{eq:prop_newton_update_aec} by an independent \ac{BNLMS} in each channel (cf.~Sec.~\ref{sec:par_update}). We conclude from Tab.~\ref{tab:perfEvalTab} that the exploitation of both algorithmic parts, i.e., echo canceller and beamformer, significantly outperforms the individual approaches (cf. LS AEC and IVE). When comparing the proposed joint update to the decoupled \ac{BNLMS}+IVE approach, we observe a significantly improved \ac{SER} which results from enhanced parameter updates of both algorithmic parts (cf. $\text{ERLE}_{\text{aec}}$ and $\text{ERLE}_{\text{bf}}$ in Tab.~\ref{tab:perfEvalTab}). Besides an improved echo attenuation, the proposed joint parameter update achieves also slightly higher interference cancellation in comparison to the individual approach. We attribute this to the higher echo attenuation provided by the echo canceller, which allows the beamformer to focus on the interferer suppression.

\begin{table}[t]
	\vspace*{-0.55cm}
	\caption{Average performance of the proposed joint \ac{AEC} and \ac{BSE} algorithm in comparison to various baselines with the best values being typeset bold.}
	\vspace*{-.3cm}
	\setlength{\tabcolsep}{4.5pt}
	\begin{center}
		\begin{tabular}{l || c c c | c c }
			Algorithm  		& {SIR} & {SER} &  {SIER} & $\text{ERLE}_{\text{aec}}$ & $\text{ERLE}_{\text{bf}}$ \\
			\midrule
			Unprocessed		& $4.78$ & $\hphantom{0}7.39$ &  $2.80$ & $0.00$ & $\hphantom{0}0.00$ \\
			\midrule
			LS AEC			& $4.78$ & $17.66$ & $4.65$ & $10.27$ & $10.27$ \\
			IVE				& $6.41$ & $10.01$ & $4.25$ & $\hphantom{0}0.00$ & $\hphantom{0}6.97$ \\
			\midrule
			BNLMS+IVE  		& $6.75$ & $17.73$ &  $6.25$ & $10.27$ & $14.40$ \\
			Joint AEC+IVE 	& $\mathbf{7.02}$ & $\mathbf{19.78}$ &  $\mathbf{6.73}$ & $\mathbf{11.39}$ & $\mathbf{16.36}$ \\		
		\end{tabular} 
	\end{center}
	\label{tab:perfEvalTab}
		\vspace{-0.3cm} 
\end{table}
%
%
%

\section{Conclusion}
\label{sec:conclusion}
In this paper, we described a computationally efficient and fast-converging joint \ac{AEC} and \ac{BSE} algorithm which improves the \ac{AEC} performance in comparison to the independent application of both algorithmic components. As future research we plan to investigate convolutive \ac{AEC} models and the incorporation of spatial prior knowledge about the \ac{SOI} \cite{9109749}.

%
%
%
%
%
%
\bibliographystyle{IEEEbib}
{\small
\bibliography{refs}

\begin{thebibliography}{10}

\bibitem{9413457}
K.~Sridhar et~al.,
\newblock ``{ICASSP} 2021 {A}coustic {E}cho {C}ancellation {C}hallenge:
  {D}atasets, {T}esting {F}ramework, and {R}esults,''
\newblock in {\em Int. Conf. Acoust., Speech, Signal Process.}, Toronto, CA,
  June 2021, pp. 151--155.

\bibitem{Kellermann1997StrategiesFC}
W.~Kellermann,
\newblock ``Strategies for combining acoustic echo cancellation and adaptive
  beamforming microphone arrays,''
\newblock in {\em Int. Conf. Acoust., Speech, Signal Process.}, Munich, DE,
  April 1997, pp. 219--222.

\bibitem{haensler2004acoustic}
E.~H{\"a}nsler and G.~Schmidt,
\newblock {\em Acoustic {E}cho and {N}oise {C}ontrol: {A} practical
  {A}pproach},
\newblock Wiley-Interscience, NJ, USA, 2004.

\bibitem{ENZNER20061140}
G.~Enzner and P.~Vary,
\newblock ``Frequency-domain adaptive {K}alman filter for acoustic echo control
  in hands-free telephones,''
\newblock {\em Signal Process.}, vol. 86, no. 6, pp. 1140--1156, 2006.

\bibitem{luis_valero_2019}
M.~L. Valero and E.~A.~P. Habets,
\newblock ``Low-{Complexity} {Multi}-{Microphone} {Acoustic} {Echo} {Control}
  in the {Short}-{Time} {Fourier} {Transform} {Domain},''
\newblock {\em IEEE Audio, Speech, Language Process.}, vol. 27, no. 3, pp.
  595--609, 2019.

\bibitem{9414868}
Mhd. Halimeh et~al.,
\newblock ``Combining {A}daptive {F}iltering and {C}omplex-{V}alued {D}eep
  {P}ostfiltering for {A}coustic {E}cho {C}ancellation,''
\newblock in {\em Int. Conf. Acoust., Speech, Signal Process.}, Toronto, CA,
  June 2021, pp. 121--125.

\bibitem{kellermann_acoustic_2001}
W.~Kellermann,
\newblock ``Acoustic {Echo} {Cancellation} for {Beamforming} {Microphone}
  {Arrays},''
\newblock in {\em Microphone {Arrays}: {Signal} {Processing} {Techniques} and
  {Applications}}, M.~Brandstein and D.~Ward, Eds., pp. 281--306. Springer
  Berlin Heidelberg, Berlin, Heidelberg, 2001.

\bibitem{Enzner2014AcousticEC}
G.~Enzner et~al.,
\newblock ``Acoustic {E}cho {C}ontrol,''
\newblock {\em Academic Press Library in Signal Process.}, vol. 4, pp.
  807--877, 2014.

\bibitem{park_state-space_2019}
J.~Park and J.-H. Chang,
\newblock ``State-{Space} {Microphone} {Array} {Nonlinear} {Acoustic} {Echo}
  {Cancellation} {Using} {Multi}-{Microphone} {Near}-{End} {Speech}
  {Covariance},''
\newblock {\em IEEE Audio, Speech, Language Process.}, vol. 27, no. 10, pp.
  1520--1534, Oct. 2019.

\bibitem{cohen_online_2021}
N.~Cohen et~al.,
\newblock ``An online algorithm for echo cancellation, dereverberation and
  noise reduction based on a {Kalman}-{EM} {Method},''
\newblock {\em Eurasip J. Audio Speech Music Process.}, vol. 2021, no. 1, pp.
  33, 2021.

\bibitem{9616295}
T.~Haubner et~al.,
\newblock ``A {S}ynergistic {K}alman- and {D}eep {P}ostfiltering {A}pproach to
  {A}coustic {E}cho {C}ancellation,''
\newblock in {\em European Signal Process. Conf.}, Dublin, IR, Aug. 2021.

\bibitem{9414180}
T.~Haubner et~al.,
\newblock ``Noise-{R}obust {A}daptation {C}ontrol for {S}upervised {A}coustic
  {S}ystem {I}dentification {E}xploiting a {N}oise {D}ictionary,''
\newblock in {\em Int. Conf. Acoust., Speech, Signal Process.}, Toronto, CA,
  June 2021, pp. 945--949.

\bibitem{9747334}
T.~Haubner et~al.,
\newblock ``{E}nd-{T}o-{E}nd {D}eep {L}earning-{B}ased {A}daptation {C}ontrol
  for {F}requency-{D}omain {A}daptive {S}ystem {I}dentification,''
\newblock in {\em Int. Conf. Acoust., Speech, Signal Process.}, Singapore, SG,
  May 2022.

\bibitem{comon1994independent}
P.~Comon,
\newblock ``Independent component analysis, a new concept?,''
\newblock {\em Signal {P}proc.}, vol. 36, no. 3, pp. 287--314, 1994.

\bibitem{nesta_batch_online}
F.~Nesta et~al.,
\newblock ``Batch-online semi-blind source separation applied to multi-channel
  acoustic echo cancellation,''
\newblock {\em IEEE Trans. Audio, Speech, Lang. Process.}, vol. 19, no. 3, pp.
  583--599, 2011.

\bibitem{gunther_learning_2012}
J.~Gunther,
\newblock ``Learning {Echo} {Paths} {During} {Continuous} {Double}-{Talk}
  {Using} {Semi}-{Blind} {Source} {Separation},''
\newblock {\em IEEE Audio, Speech, Language Process.}, vol. 20, no. 2, pp.
  646--660, 2012.

\bibitem{9414623}
Z.~Wang et~al.,
\newblock ``Weighted {Recursive} {Least} {Square} {Filter} and {Neural}
  {Network} {Based} {Residual} {Echo} {Suppression} for the
  {AEC}-{Challenge},''
\newblock in {\em Int. Conf. Acoust., Speech, Signal Process.}, Toronto, CA,
  June 2021, pp. 141--145.

\bibitem{9357975}
G.~Cheng et~al.,
\newblock ``Semi-blind source separation for nonlinear acoustic echo
  cancellation,''
\newblock {\em IEEE Signal Process. Lett.}, vol. 28, pp. 474--478, 2021.

\bibitem{4538677}
H.~Buchner and W.~Kellermann,
\newblock ``A fundamental relation between blind and supervised adaptive
  filtering illustrated for blind source separation and acoustic echo
  cancellation,''
\newblock in {\em Hands-Free Speech Comm and Microphone Arrays (HSCMA)},
  Trento, IT, May 2008, pp. 17--20.

\bibitem{na21_interspeech}
Y.~Na et~al.,
\newblock ``{Joint Online Multichannel Acoustic Echo Cancellation, Speech
  Dereverberation and Source Separation},''
\newblock in {\em Interspeech}, Brno, CZ, Sept. 2021, pp. 1144--1148.

\bibitem{9657518}
C.~Boeddeker et~al.,
\newblock ``A comparison and combination of unsupervised blind source
  separation techniques,''
\newblock in {\em ITG Conf. on Speech Comm.}, Kiel, DE, Sept. 2021.

\bibitem{grad_alg_quest_conv}
Z.~Koldovský and P.~Tichavský,
\newblock ``Gradient algorithms for complex non-gaussian independent
  component/vector extraction, question of convergence,''
\newblock {\em IEEE Trans. Signal Process.}, vol. 67, no. 4, pp. 1050--1064,
  2019.

\bibitem{van_trees}
H.~L.~Van Trees,
\newblock {\em Optimum Array Processing: Part {IV} of Detection, Estimation,
  and Modulation Theory},
\newblock Wiley, 2002.

\bibitem{cois1994performance}
J.-F. Cardoso,
\newblock ``On the performance of orthogonal source separation algorithms,''
\newblock in {\em European Signal Process. Conf.}, Edinburgh, UK, 1994,
  vol.~94, pp. 776--779.

\bibitem{8081389}
Z.~Koldovský et~al.,
\newblock ``Orthogonally {C}onstrained {I}ndependent {C}omponent {E}xtraction:
  {B}lind {MPDR} {B}eamforming,''
\newblock in {\em European Signal Processing Conf.}, Kos Island, GR, Aug. 2017,
  pp. 1155--1159.

\bibitem{complex_ad_filt}
H.~Li and T.~Adali,
\newblock ``Complex-valued adaptive signal processing using nonlinear
  functions,''
\newblock {\em EURASIP J. Adv. Signal Process.}, vol. 2008, Jan. 2008.

\bibitem{9387552}
Z.~Koldovský et~al.,
\newblock ``Dynamic independent component/vector analysis: Time-variant linear
  mixtures separable by time-invariant beamformers,''
\newblock {\em IEEE Trans. Signal Process.}, vol. 69, pp. 2158--2173, 2021.

\bibitem{fica}
A.~Hyv\"{a}rinen and E.~Oja,
\newblock ``A fast fixed-point algorithm for independent component analysis,''
\newblock {\em Neural Comput.}, vol. 9, no. 7, pp. 1483–1492, Oct. 1997.

\bibitem{matsuoka2002minimal}
K.~Matsuoka,
\newblock ``Minimal distortion principle for blind source separation,''
\newblock in {\em SICE Annual Conf.}, Osaka, JP, Aug. 2002, pp. 2138--2143.

\bibitem{habets2010room}
E.~A.~P. Habets,
\newblock ``Room {I}mpulse {R}esponse {G}enerator,''
\newblock Tech. {R}ep., Technische Universiteit Eindhoven, Sept. 2010.

\bibitem{libri_speech}
V.~{Panayotov} et~al.,
\newblock ``Librispeech: An {ASR} corpus based on public domain audio books,''
\newblock in {\em Int. Conf. Acoust., Speech, Signal Process.}, Brisbane, AU,
  Apr. 2015, pp. 5206--5210.

\bibitem{9109749}
A.~Brendel et~al.,
\newblock ``A {U}nified {P}robabilistic {V}iew on {S}patially {I}nformed
  {S}ource {S}eparation and {E}xtraction {B}ased on {I}ndependent {V}ector
  {A}nalysis,''
\newblock {\em IEEE Trans. Signal Process.}, vol. 68, pp. 3545--3558, 2020.

\end{thebibliography}
}

\end{document}